\documentclass[aps,pre,twocolumn,titlepage,nofootinbib]{revtex4}
\usepackage{graphicx}
\usepackage{color}
\usepackage{dcolumn}
\usepackage{latexsym}

\usepackage[normalem]{ulem}
\usepackage{hyperref,amssymb}
\usepackage{url}
\usepackage{color,soul}
\usepackage{graphicx}
\usepackage{verbatim}
\usepackage{multirow}
\usepackage{amsmath}
\newcommand{\beq}{\begin{eqnarray}}
\newcommand{\eeq}{\end{eqnarray}}
\usepackage{mathrsfs}
\usepackage{float,soul}
\usepackage[dvipsnames]{xcolor}
\usepackage{mathtools}
\usepackage{slashed}
\usepackage{physics}	
\usepackage{graphicx}   
\usepackage{epstopdf}
\usepackage{subfigure}  
\usepackage{hyperref}   
\usepackage{bbold}
\usepackage{wasysym}
\usepackage{feynmp}
\usepackage{hyperref}
\hypersetup{colorlinks,
}

\usepackage[latin1]{inputenc}

\begin{document}

\title{\Large Clustering of negative topological charge precedes plastic failure in 3D glasses}
\author{Arabinda Bera$^{1}$}

\author{Matteo Baggioli$^{2}$}
\email{b.matteo@sjtu.edu.cn}
\author{Timothy C. Petersen$^{3}$}
\author{Timothy W. Sirk$^{4}$}
\author{Amelia C. Y. Liu$^{5}$}
\email{amelia.liu@monash.edu}
\author{Alessio Zaccone$^{1}$}
\email{alessio.zaccone@unimi.it}
\address{$^1$Department of Physics ``A. Pontremoli", University of Milan, via Celoria 16, 20133 Milan, Italy}
\address{$^2$Wilczek Quantum Center, School of Physics and Astronomy, Shanghai Jiao Tong University, Shanghai 200240, China}
\address{$^3$Monash Centre for Electron Microscopy, Monash University, Clayton, 3800, Victoria, Australia \& School of Physics and Astronomy, Monash University, Clayton, 3800, Victoria, Australia}
\address{$^4$US DEVCOM Army Research Laboratory, Maryland, United States}
\address{$^5$School of Physics and Astronomy, Monash University, Clayton, 3800, Victoria, Australia}

\begin{abstract}
The deformation mechanism in amorphous solids subjected to external shear remains poorly understood because of the absence of well-defined topological defects mediating the plastic deformation. The notion of soft spots has emerged as a useful tool to characterize the onset of irreversible rearrangements and plastic flow, but these entities are not clearly defined in terms of geometry and topology. In this study, we unveil the phenomenology of recently discovered, precisely defined topological defects governing the microscopic mechanical and yielding behavior of a model 3D glass under shear deformation. We identify the existence of vortex-like and anti-vortex-like topological defects within the 3D non-affine displacement field. The number density of these defects exhibits a significant anti-correlation with the plastic events, with defect proliferation-annihilation cycles matching the alternation of elastic-like segments and catastrophic plastic drops, respectively. Furthermore, we observe collective annihilation of these point-like defects via plastic events, with large local topological charge fluctuations in the vicinity of regions that feature strong non-affine displacements. We reveal that plastic yielding is driven by several large sized clusters of net negative topological charge, the massive annihilation of which triggers the onset of plastic flow. These findings suggest a geometric and topological characterization of soft spots and pave the way for the mechanistic understanding of topological defects as mediators of plastic deformation in glassy materials.
\end{abstract}

\maketitle


While plastic deformations in crystals are by now well understood in terms of dislocation dynamics, a concept originally proposed by Taylor \cite{Taylor}, identifying the physical mechanism of plasticity in amorphous solids, such as glasses, is more challenging \cite{WANG201581}. 
Despite several attempts to generalize the concept of topological defects and dislocation-like structures in amorphous solids \cite{10.1063/1.1662243,PhysRevLett.43.1517,doi:10.1080/01418618108235816,doi:10.1080/01418618008243894}, not much progress in this direction has been achieved so far, and this approach is still considered by many as hopeless. The deformed structure is indistinguishable from the disordered reference structure \cite{ALEXANDER199865}. Other existing viewpoints on this matter, such as shear transformation zone (STZ) theory \cite{PhysRevE.57.7192}, 2D melting-type scenarios \cite{Fraggedakis} or elastoplastic models \cite{RevModPhys.90.045006}, assume the existence of soft spots linked to some unspecified defects, but they are purely phenomenological and agnostic about the precise definition of the defect. In particular, soft spots, as discussed in the literature \cite{PhysRevLett.107.108302}, lack a quantitative geometric characterization or measure to identify them, in contrast to topological defects in crystalline matter and liquid crystals \cite{Fumeron2023}. At the same time, novel ideas to identify defects in amorphous solids based on quasi-localized modes \cite{10.1063/5.0069477}, geometric charges \cite{PhysRevE.104.024904} or string-like defects \cite{PhysRevB.91.094102,ZHOU2023118701} are not mature enough; consequently, structural indicators \cite{PhysRevMaterials.4.113609} still remain the best available option to predict plasticity in glasses.

Recently, substantial progress has been achieved by looking for defects beyond the paradigm of the static structure. Following this idea, well-defined topological defects have been discovered in the shear deformation of simulated polymer glasses \cite{baggioli2021}. These defects were identified with singularities in the displacement field, which emerge in the non-affine part $\bf{u}$ of the total displacement field $\bf{u_t}=\bf{u_A}+\bf{u}$, where $\bf{u_A}$ is the smooth affine contribution \cite{PhysRevB.66.174205,PhysRevE.72.066619,zaccone2023}. As for dislocations in ordered crystals \cite{kleinert1989gauge}, a circulation integral around such topological defects connects to the incompatibility of the strain field, \textit{i.e.} the curl of the curl of the strain tensor does not vanish. This is, in turn, mathematically equivalent to a Bianchi identity for the displacement field which is violated topologically due to the presence of a finite Burgers vector, and to the explicit breaking of an emergent higher-form symmetry \cite{baggioli2022}.

Shortly thereafter, further evidence of the existence of well-defined topological defects in glasses came from the analysis of eigenvectors in 2D simulated glasses \cite{wu2023,Baggioli2023}. It was found that, even without applying any deformation to the 2D glass, a circulation integral of the angle between the $y$ and $x$ components of the eigenvectors of the Hessian matrix yields vortex-like defects with quantized $+1$ charge, and anti-vortex-like defects with quantized $-1$ charge. Moreover, spatial correlation between the negative topological charges and the soft regions most inclined to plastic deformation was found. Interestingly, the dynamics of vortex-like defects in the displacement field have been already connected in the past to the formation of shear transformation zones \cite{PhysRevLett.119.195503,SOPU2023170585}, confirming their importance for the description of plasticity in amorphous solids.

Almost at the same time with the current paper, a preprint \cite{desmarchelier2024topological} has been posted, where the concept of topological defects has been used to unambiguously identify shear transformation zones in 2D amorphous solids, providing a direct connection between individual topological defects with negative charge and local structural rearrangements responsible for plasticity. Topological defects can be modelled in a continuum fashion using geometry, and in particular relating dislocations to torsion and disclinations to curvature \cite{kleinert1989gauge,RevModPhys.80.61}. Likewise, point defects can be described using a geometric approach \cite{kupferman2018,yavari_new}. The relationship between these geometric charges and the vortex-like defects considered in this article remains unclear. Our work investigates the connection between topological defects in the non-affine displacement field, as examined in recent studies \cite{wu2023, desmarchelier2024topological}, and the mechanical properties of a three dimensional polymer glass system under shear deformation. 

In this study, we establish a protocol for identifying quantized ($\pm 1$) topological defects in the displacement field of model 3D glasses undergoing shear deformation. Through defect analysis of quasi-2D slices within the shear plane, we demonstrate a correlation between the population of topological defects and plastic drops in the stress-strain curve. Further dissection of these 2D slices reveals microscopic details on local topological charge and associated non-affinity. Notably, fluctuations in topological charge are observed in proximity to regions exhibiting strong non-affine displacements, indicating the emergence of soft spots. Systematic analysis of various 2D slices enables the construction of a three-dimensional representation of topological charges originating from defects, highlighting the drastic annihilation of such defects during yielding -- a ``topological avalanche'' \cite{https://doi.org/10.1103/RevModPhys.76.471}.

\section*{Results}
Here we prepare a polymer glass sample at $T=0$ implemented using the Kremer-Grest model \cite{grest1986}. The system is then subjected to an athermal quasi-static (AQS) shearing protocol involving affine transformation by small strain increments of $\delta \gamma_{xz} =0.001$ in the $xz$ plane following energy minimization. More details on the interaction potential and the deformation protocol of the simulations are provided in the method section.

During the deformation, we determine the total displacement field ${\bf{u_t}}$ by taking the difference between two successive snapshots separated by $\delta \gamma_{xz} = 0.001$. The affine displacement field here is calculated as ${\bf{u_A}}=\delta \gamma_{xz}z_{\text{eff}} \hat{x}$, where $z_{\rm{eff}}$ represents the effective distance from the base of the box along the $z$-direction. 

The non-affine field is defined as ${\bf{u}}={\bf{u_t}}-{\bf{u_A}}$ \cite{zaccone2023}, and can be expressed as ${\bf{u}} = \bf{u_{\parallel}}+\bf{u_{\perp}}$, where $\bf{u_{\parallel}}$ is the vector in the $xz$ (shear) plane, and $\bf{u_{\perp}}$ is the vector along the $y$-direction. The average magnitude of non-affine displacement in three dimensions is computed as $u_{\rm{mag}}^{3D} = \frac{1}{N}\sum_{i=1}^{N}|\bf{u}_i|$. This average is performed over the total number of monomers/particles ($N=10000$) in the system. As the shear is applied along the $x$-direction to deform the $xz$ (shear) plane, we compute the average non-affine displacement in the $xz$ plane as  $u_{\rm{mag}}=\langle \frac{1}{N_s}\sum_{i=1}^{N_s}|\bf{u_{{\parallel},i}}| \rangle$, where $\langle ... \rangle$ denotes the average over all slices along the $y$ direction, and $N_s$ represents the number of particles in each two-dimensional (2D) slice. For our analysis, we divided the 3D box into numerous two-dimensional slices of thickness $2\sigma$~($\sigma$ is the monomer diameter) along the $y$ direction.

\begin{figure}[h]
\centering
\includegraphics[width=\linewidth]{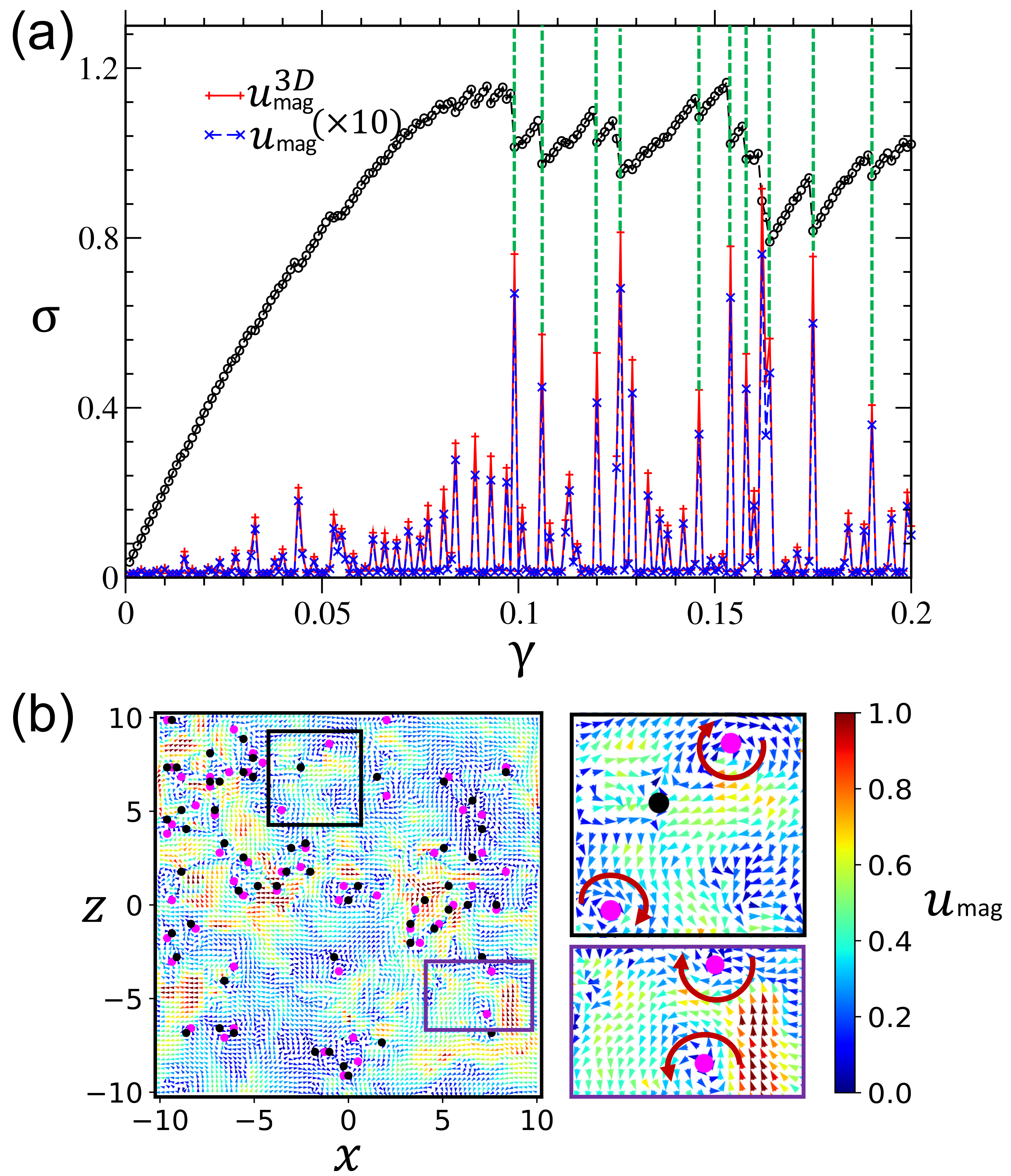}
\caption{\textbf{(a)} The stress versus strain curve is plotted, accompanied by red lines representing the average magnitudes of non-affine displacement in three dimensions ($u_{\rm{mag}}^{3D}$) and blue lines illustrating the average magnitude of the non-affine field in the $xz$ plane ($u_{\rm{mag}}$). The green dashed lines are indicative of the ``major plastic events''. \textbf{(b)} Normalized non-affine displacement field presented for a 2D slice within the $xz$ plane at $y=0$, with $\gamma=0.098$. Circles filled in magenta and black represent topological defects with winding numbers $q=+1$ and $-1$, respectively. Enlarged views of specific regions (outlined by the rectangular boxes) emphasize the chirality of the defects (described by the red arcs). The color bar illustrates the magnitude of non-affine displacement.}
\label{fig1}
\end{figure}

In Fig.\ref{fig1}, we present the stress ($\sigma$) versus strain ($\gamma$) curve. Additionally, we plot the average magnitudes of the non-affine displacement, denoted as $u_{\rm{mag}}^{3D}$ and $u_{\rm{mag}}$, represented by red and blue lines, respectively. The stress drop is notably identifiable by the peaks in the non-affine displacement values. These peaks in $u_{\rm{mag}}^{3D}$ and $u_{\rm{mag}}$ are evidently commensurate with the occurrence of plastic events (PEs). We have highlighted several of these PEs using green dashed lines in Fig.\ref{fig1}. Remarkably, we observe that $u_{\rm{mag}}$, which disregards the $y$-component of the field, can accurately identify the locations of PEs. This observation implies that the characteristics of displacement field remain consistent when we exclude the $y$-component (orthogonal to the shear plane). However, the same behavior is observed even when we exclude the $x$ or $z$ component of the non-affine field, as proven explicitly in the Supplementary Information. This suggests that the peaks in $u_{\rm{mag}}$ occur due to the high values of the non-affine field magnitude for a small fraction of all particles. All three components of the non-affine displacement for these fraction of particles have approximately of the same order of magnitude. This is why discarding any one component does not alter the overall characteristics of the $u_{\rm{mag}}$ versus $\gamma$ curve.

Moreover, we have confirmed the convergence of our numerical analysis by considering also smaller strain steps, $\delta\gamma=10^{-4}$, and obtained analogous results. As a direct proof, we compare the magnitude of non-affine displacement for different strain steps in the Supplementary Information.

We investigate point-like topological defects by analyzing the non-affine displacement fields within various 2D slices orthogonal to the $y$ direction. We construct circulation integrals along closed loops $\mathcal{L}$ and calculate the winding number $q$ as:
\begin{equation}\label{topo_charge}
q = \frac{1}{2\pi} \oint_{\mathcal{L}} d\theta = \frac{1}{2\pi} \oint_{\mathcal{L}} \vec{\nabla} \theta \cdot d\vec{\ell},
\end{equation}
where $\theta$ represents the phase angle of the non-affine field, defined as $\theta = \arctan(u^{z}/u^{x})$ with reference to the Cartesian $z$ and $x$ components of the non-affine displacement field in the shear plane. Furthermore, $d\vec{\ell}$ denotes the line elements along the closed loop $\mathcal{L}$. The winding number $q\in \mathbb{Z}$ is referred to as the topological charge. 

In 2D, a point defect carrying a winding number of $q=+1$ is recognized as a vortex, while a defect with $q=-1$ is identified as an anti-vortex. We systematically discerned these point defects across various 2D slices within the $xz$ plane. As depicted in Fig.\ref{fig1}(b), we present the interpolated non-affine displacement field with a color bar indicating the scaled magnitude, within a 2D slice at $y=0$ for $\gamma=0.098$. The existence of topological defects with winding numbers $q=+1$ and $q=-1$ is marked by magenta and black-filled circles, respectively. Notably, two rectangular regions are selectively magnified to elucidate the chirality of vortices ($q=+1$). At the frustrated interface between two vortices of identical phase chirality (direction of swirls), an anti-vortex ($q=-1$) is observed. Conversely, vortices with opposing phase chiralities exhibit a stable interface, as portrayed in the enlarged sections of Fig. \ref{fig1}(b). This intriguing pattern bears resemblance to topological defects discerned in the eigenvectors field of a recently investigated 2D glass \cite{wu2023}. As shown in \cite{desmarchelier2024topological}, the displacement and stress distribution of $-1$ defects strongly resemble those of an Eshelby inclusion with quadrupolar structure, \textit{i.e.} a shear transformation zone. The non-affine displacement field incorporates contributions from the entire spectrum of eigenmodes of the Hessian matrix. In the work by Tanguy \cite{PhysRevB.66.174205}, it is reported that the structure of the eigenvectors from the Hessian matrix and that from the nonaffine displacement field strongly correlate in space for low eigenfrequencies. Since plastic events are typically dominated by low-frequency modes, it can be expected that analyses based on these low-frequency modes and the displacement field yield consistent results. However, this aspect is beyond the scope of our current study.\\

\begin{figure}[h]                      
\centering
\includegraphics[width=\linewidth]{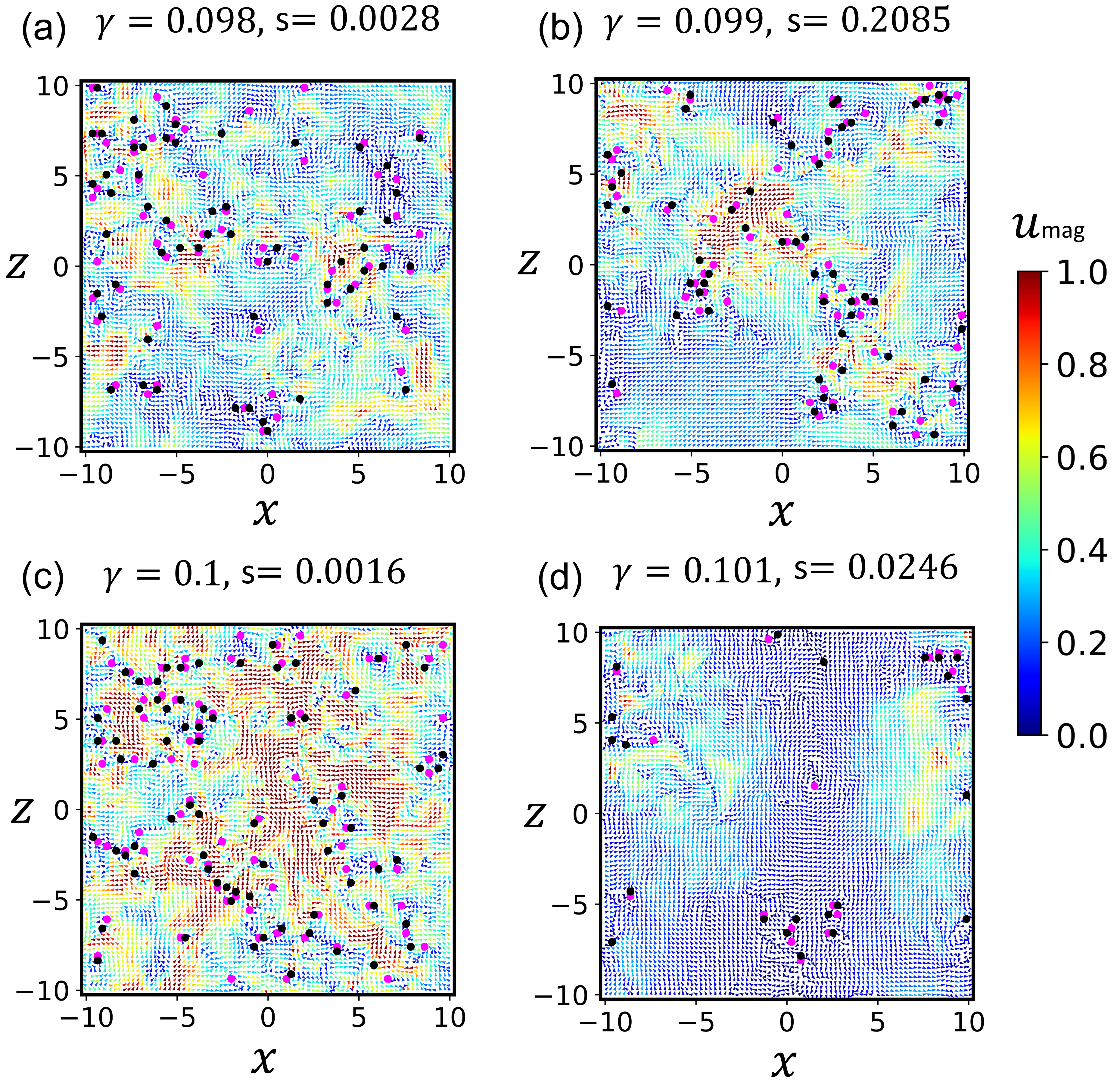}
\caption{\textbf{(a)-(d)}~Normalized displacement field for four consecutive $\gamma$ values within the 2D slice at $y=0$. Topological defects are represented by circles filled by magenta ($q=+1$) and black ($q=-1$). The color bar (scaled by the factor $s$) indicates the magnitude of the displacement field. Note that the range of the displacements varies by two orders of magnitudes and the displacements are extremely large around yielding (panel (b)).}
\label{fig2}
\end{figure}
       
In Fig.\ref{fig2}, we present the locations of point-defects within the $xz$ plane non-affine displacement field at $y=0$ for four consecutive $\gamma$ values that straddle the yielding point. These snapshots reveal a significant transformation in the structure of defects. We compute the total number of these point defects in the non-affine displacement field across various $xz$ planes by varying $y$ and calculate the average number of defects $N_d$ per slice. In order to establish a connection between plastic events and the emergence of point-like topological defects, we compute $N_d$ with increasing strain $\gamma$. In our analysis we interpolate the displacement field, ignoring the periodic boundary conditions, and additionally, we consider two dimensional slices of a 3D system. Consequently, the number of vortices and anti-vortices is not completely equal in any given interpolated 2D displacement field. However, when averaging the defect numbers over several 2D slices, the number of defects with opposite charge is approximately equal; leading to near-zero total topological charge, or approximate conservation of topological charge.\\

\begin{figure}[h]  
\centering
\includegraphics[width=\linewidth]{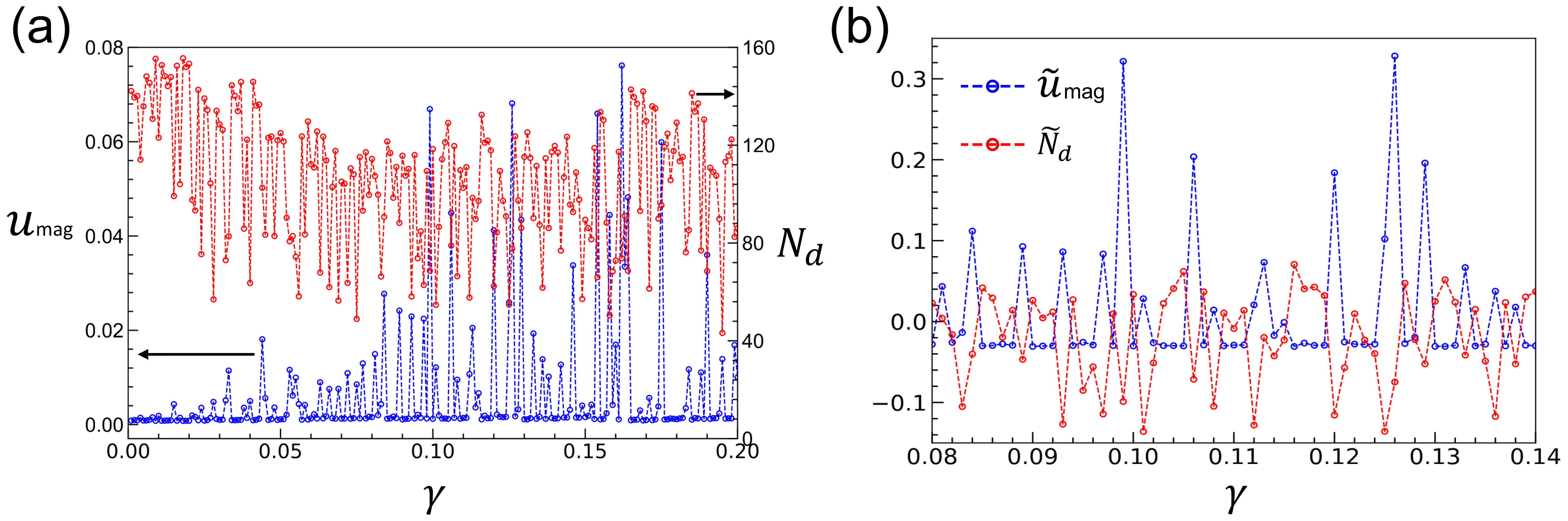}
\caption{\textbf{(a)} The magnitude of non-affine displacement field $u_{\rm{mag}}$ (left) and the average defect number $N_d$ (right) are shown for the entire range of $\gamma$ values. \textbf{(b)} The quantities, $\tilde{u}_{\rm{mag}}$ and $\tilde{N}_d$ are presented for a limited range of $\gamma$ for better visualization.}
\label{fig3}
\end{figure}

In Fig.\ref{fig3}(a), we depict the average magnitude of the displacement field ($u_{\rm{mag}}$) and the average number of defects ($N_d$) per slice across the entire range of $\gamma$. To establish a correlation, we calculate the dimensionless quantities $\tilde{u}_{\rm{mag}}$ and $\tilde{N}_d$, defined as $\tilde{u}_{\rm{mag}}=({u}_{\rm{mag}}-\bar{u}_{\rm{mag}})/{{\sigma}_{u_{\rm{mag}}}}$ and $\tilde{N}_d=({N}_d-\bar{N}_d)/{{\sigma}_{N_d}}$. Here $\bar{u}_{\rm{mag}}$ and $\bar{N}_d$ denote the average values of $u_{\rm{mag}}$ and $N_d$ over the range of strain $\gamma \in [0,0.2]$, respectively. The quantities ${{\sigma}^2_{u_{\rm{mag}}}}={\sum_{\{\gamma\}}({u}_{\rm{mag}}-\bar{u}_{\rm{mag}})^2}$ and ${{\sigma}^2_{N_d}}={\sum_{\{\gamma\}}({N}_{d}-\bar{N}_{d})^2}$ are proportional to the variances of $u_{\rm{mag}}$ and $N_{d}$, respectively. In Fig.\ref{fig3}(b), we illustrate the variation of $\tilde{u}_{\rm{mag}}$ and $\tilde{N}_d$ within the range of $\gamma$. Notably, the peaks in $u_{\rm{mag}}$ align with dips in the total number of point defects. This observation strongly suggests an anti-correlation between $u_{\rm{mag}}$ and $N_d$, implying that the accumulation of topological defects triggers plastic events through subsequent defect annihilation. This anti-correlation might be explained by the fact that topological defects occur where the displacement field is close to zero and fluctuating, and thus singular or not defined. It is thus plausible that these defects anti-correlate with plasticity, as zeros in the displacement field are less common at strain steps pertaining to large plastic events, concomitant with large and coordinated displacements over extensive regions.\\

As apparent from Fig.\ref{fig3}(a), the topological defects are of transient nature. In the Supplementary Information, we have analyzed this aspect further by extending the analysis to smaller strain step, $\delta \gamma =10^{-4}$. We observe drastic fluctuations in $N_d$ in this case as well, indicating that
the transient nature persists even for smaller $\delta \gamma$. In order to understand the origin of this transient nature, it would be useful to investigate in detail the dynamics of these defects, their lifetime and their interactions.

To explore this dependence further, we calculate the Pearson correlation coefficient (${\rm{PCC}}$) \cite{pearson1895}, given by ${\rm{PCC}}=\sum_{\{\gamma\}}\tilde{u}_{\rm{mag}} \tilde{N}_d$. The value of ${\rm{PCC}}$ falls within the range of [-1, 1], where a value close to 1 indicates a strong positive correlation, while a value near -1 suggests a strong negative correlation. Averaging over the entire range of $\gamma$, we obtain the value of ${\rm{PCC}} \approx -0.5$, indicating a marked level of anti-correlation. Interestingly, when we consider only the data points associated with plastic events, identified from the peaks of $u_{\rm{mag}}$, the value of ${\rm{PCC}}$ approaches $-1$. This suggests that there is a strong (inverse) relationship between the topological defects and plastic deformation events in the material.\\

\begin{figure}[h]  
\centering
\includegraphics[width=\linewidth]{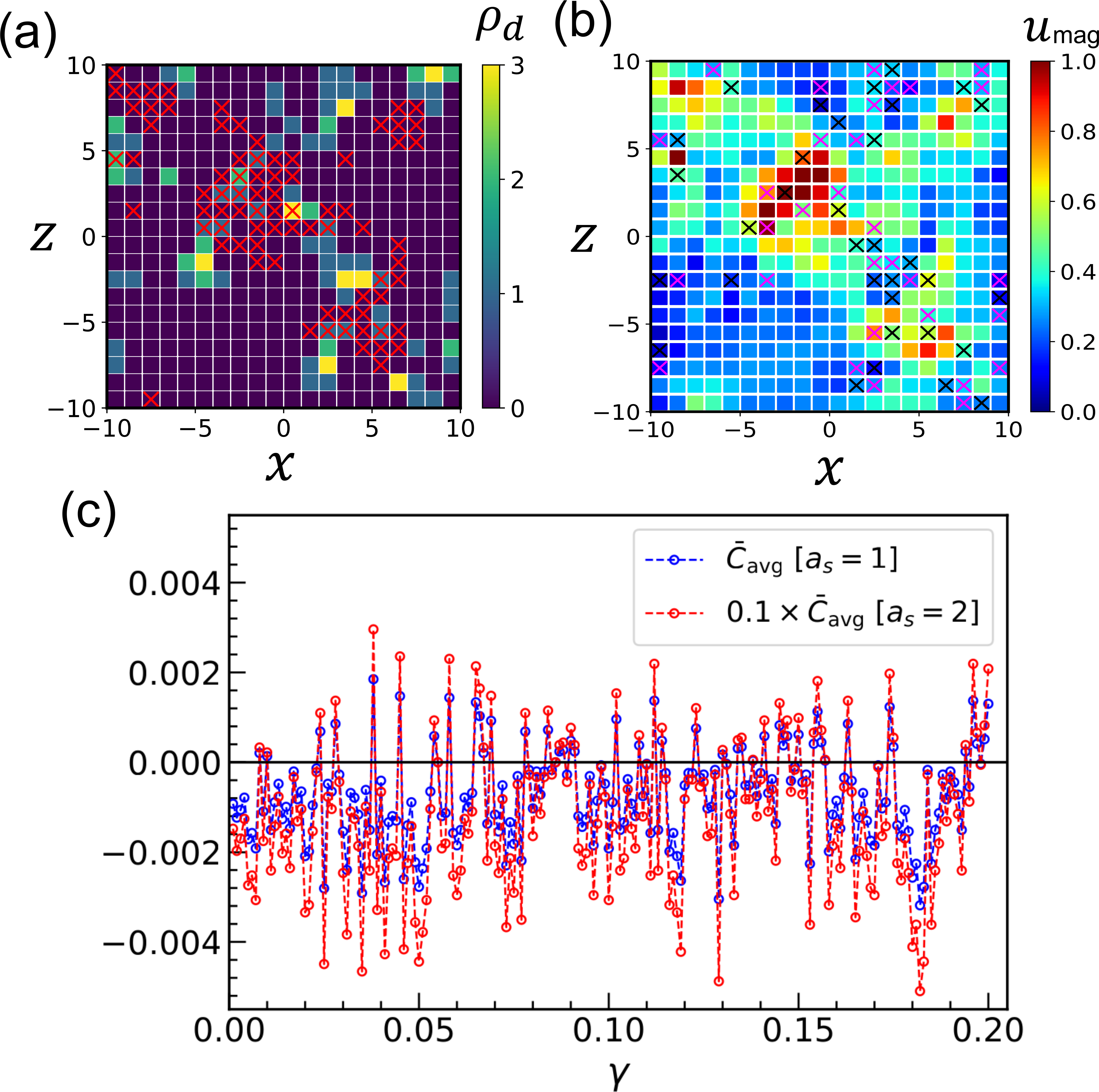}
\caption{\textbf{(a)} A 2D slice ($\gamma=0.099$~and~$y=0$) dissected into many unit area square cells is shown with the local topological number density $\rho_d$ in each cell by color bar. The cells having larger non-affine magnitude have been presented by red crosses. \textbf{(b)} The same slice, as in (a), shown by color (scaled by 0.2) which represents the local magnitude of non-affine displacement. The cells marked by crosses colored in magenta and black are populated by topological charges with positive and negative values, respectively. \textbf{(c)} The average topological charge density $\bar{C}_{\mathrm{avg}}$ per cell is shown as a function of $\gamma$ for two different coarse-grained sizes of the sub-square box $a_s \times a_s$.}
\label{fig4}
\end{figure}

\begin{figure}[h]  
\centering
\includegraphics[width=\linewidth]{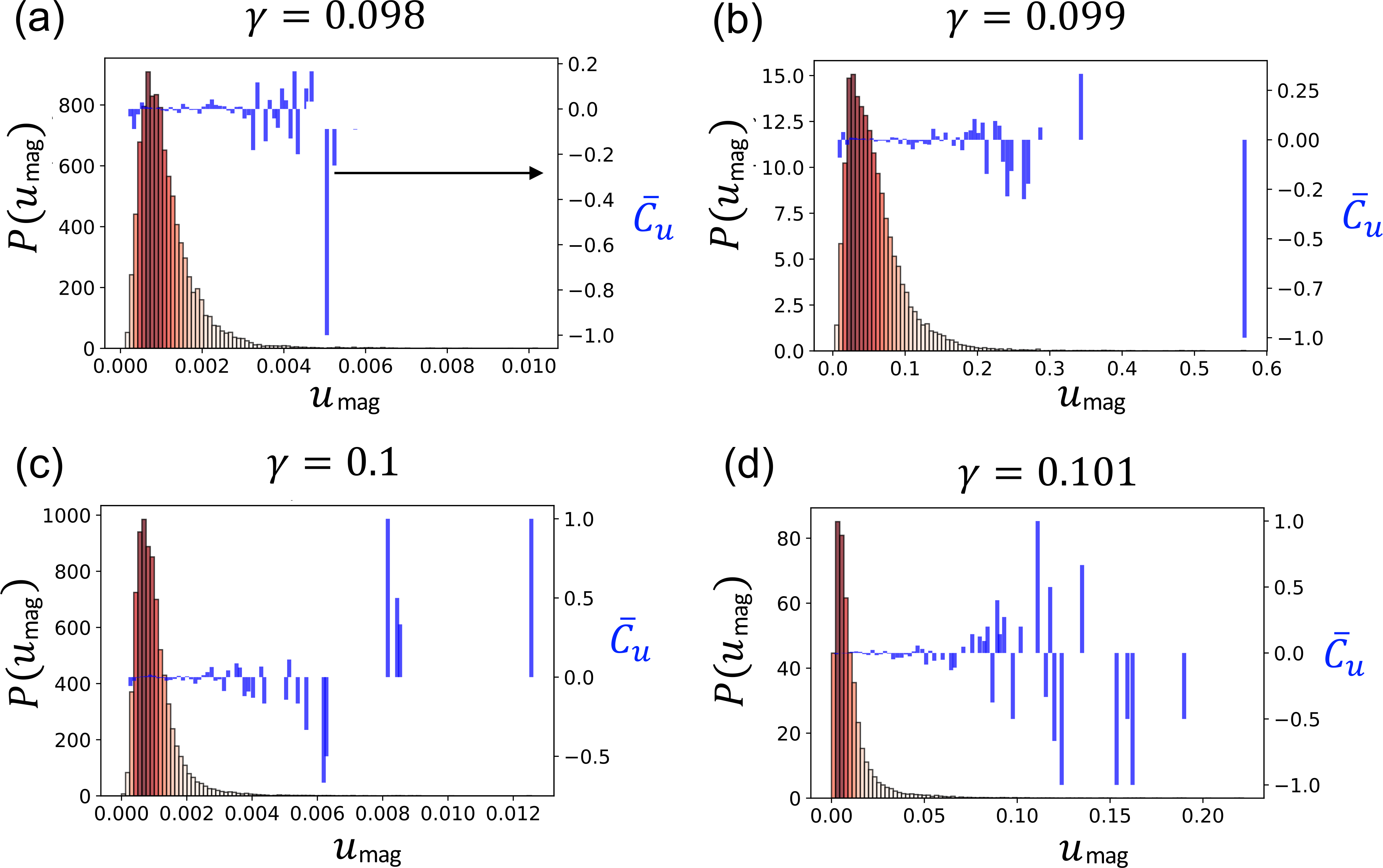}
\caption{\textbf{(a)-(d)} The distribution of $u_{\rm{mag}}$ and associated average topological charge $\bar{C}_u$ per cell are shown for several $\gamma$ values. Note that there is large variation in the magnitude of the displacements as a function of strain.
}
\label{fig5}
\end{figure}

While we have established a robust anti-correlation between plastic events and the variation in the total number of topological point defects, our primary objective is to understand the microscopic details of these plastic events and investigate the specific role played by the topological defects. Upon examining the snapshots in Figs.\ref{fig1}(b) and \ref{fig2}, we observe that topological defects tend to be positioned in close proximity to regions characterized by high non-affine displacement magnitudes. In order to investigate whether the local non-affine displacement field correlates with the presence of topological defects, we systematically partition each 2D slice into numerous unit area square cells, as illustrated in Figs.\ref{fig4}(a) and (b). Within each such cell, we compute the number density of defects $n_d$, average non-affine displacement $u_{\rm{mag}}$ and concurrently measure the local topological charge in each cell, denoted as $\bar{C}_k=\sum_{j\in \mathbb{S}_k}q_j$, where $q_j$ is the topological charge of defects present inside the $k$-th cell, $\mathbb{S}_k$. In fact, $\bar{C}_k$ essentially measures the topological charge imbalance inside each cell. The average topological charge imbalance is computed as $\bar{C}_{\mathrm{avg}}=(1/n_{\rm{cell}})\sum_{k=1}^{n_{\rm{cell}}}\bar{C}_k$, where $n_{\rm{cell}}$ is the total number of square cells. Let us emphasize that this analysis probes a mesoscopic coarse-grained length scale corresponding to the size of the (square) sub-cell of length $\sigma$ and does not reveal any local microscopic information.

In Fig.\ref{fig4}(a) and (b), we visually illustrate the partitioning of a 2D slice into multiple square cells, utilizing a color bar to represent the number density of topological defects ($n_d$) and the magnitude of the non-affine displacement field ($u_{\rm{mag}}$ within each cell), respectively. The red crosses in Fig. \ref{fig4}(a) pinpoint cells where $u_{\rm{mag}}$ is notably higher compared to other cells, while colours of the crosses within several cells in Fig.\ref{fig4}(b) demarcate the signs of the average topological charge, denoted as $\bar{C}_{\rm{avg}}$. Magenta highlights cells with $\bar{C}_k \ge 1$, and black designates cells with $\bar{C}_k \le -1$. In Fig.\ref{fig4}(c), we present the average local topological charge imbalance $\bar{C}_{\rm{avg}}$, computed across all cells and various 2D slices, while varying $\gamma$. Here, we depict the variation of $\bar{C}_{{\rm{avg}}}$ for two different square cell areas with sides $a_s=1$ and $2$. The behavior of $\bar{C}_{{\rm{avg}}}$ versus $\gamma$ remains consistent across varying coarse-grained cell sizes, except for a constant scale factor reflecting the different cell areas. Remarkably, we observe a prevailing bias in local charge imbalance towards net negative charge values, which we attribute as an underlying tendency towards plastic deformation.

In Fig.\ref{fig5}(a)-(d), we further analyze the distribution of $u_{\rm{mag}}$ and the corresponding average topological charge imbalance $\bar{C}_u$ across different $u_{\rm{mag}}$ values, where the average is performed over cells with same $u_{\rm{mag}}$. We observe much greater fluctuations in local charge imbalance at higher non-affine displacements. The pronounced charge imbalance at higher non-affine field suggests the formation of mesoscopic soft spots possessing a net topological charge, upon approaching yielding.\\

\begin{figure}[h]                               
\centering
\includegraphics[width=\linewidth]{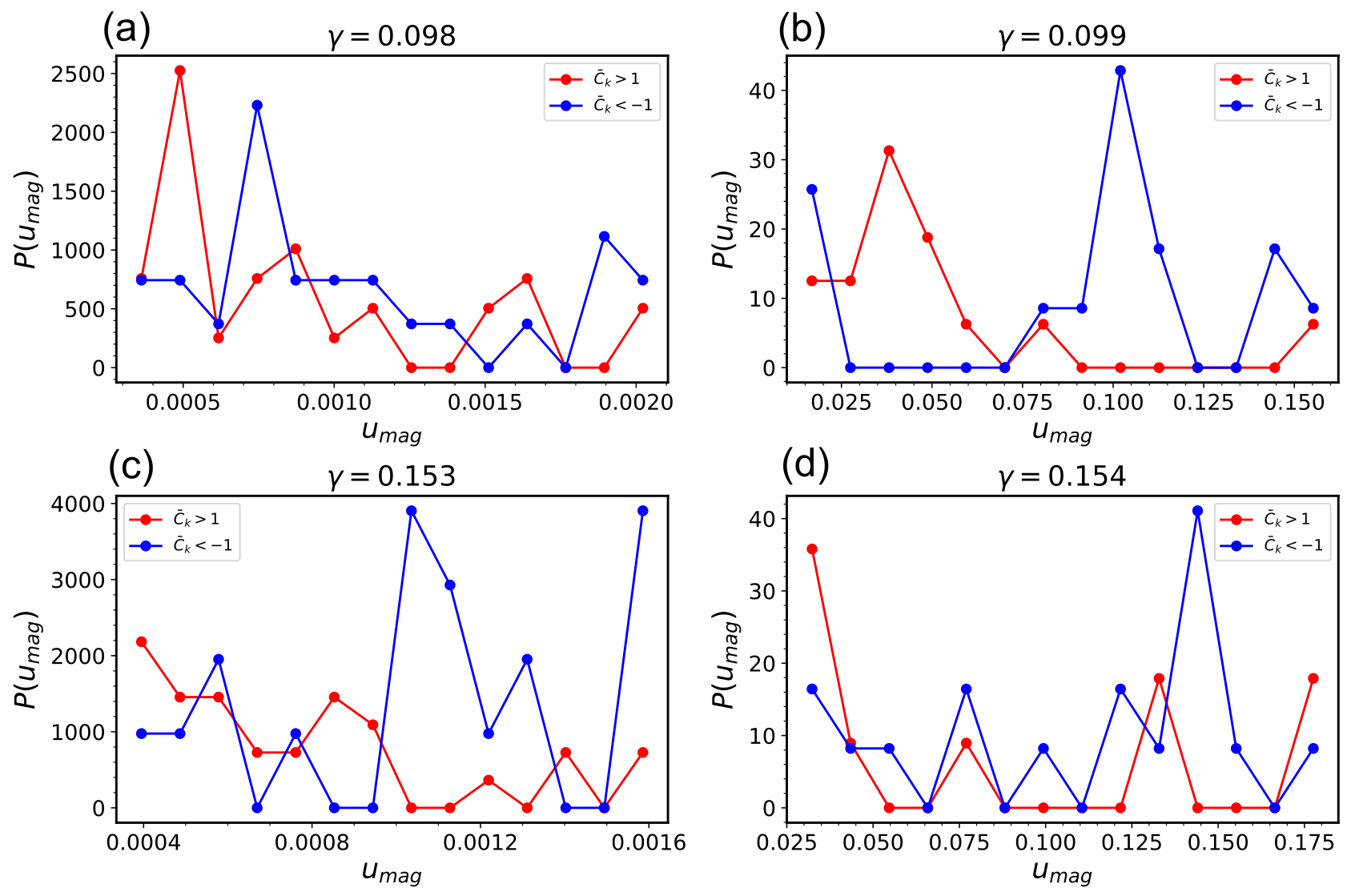}
\caption{\textbf{(a)-(d)} The normalised distribution function that measures the probability of obtaining topological charge $\bar{C}_k>1$ and $\bar{C}_k<-1$ at different local regions having average non-affine displacement magnitude ($u_{\rm{mag}}$) are shown at the vicinity of two major plastic events at $\gamma=0.099$ and $\gamma=0.154$.}
\label{fig6}
\end{figure}

To investigate the development of these soft spots, we calculate the distribution of $u_{\rm{mag}}$ for coarse-grained cells with local topological charge $\bar{C}_k>1$ or $\bar{C}_k<-1$. In Fig. \ref{fig6}, we present this distribution for positive and negative charge imbalances towards the approach of two plastic events at $\gamma=0.099$ and $\gamma=0.154$. Notably, distinct peaks in these distributions at high $u_{\rm{mag}}$ are almost always associated with a net negative charge, showing how the negative defects clusters emerge as soft spots at the onset of yielding.\\

Our analysis has primarily focused on various 2D cross-sections within the $xz$ plane along the $y$ axis. In order to extend our observations into three dimensions, we developed a generalized protocol to systematically scan the system from arbitrary directions. Positions of topological defects were identified across numerous two-dimensional slices, each oriented along a direction vector $\hat{n}$. By considering various directions $\hat{n}$ uniformly distributed on a unit sphere, we located defects within the simulation box. Similar to the dissection of 2D slices illustrated in Fig. \ref{fig4}(a) and (b), we partitioned the entire simulation box into cubical cells of unit volume. Utilizing the locations of topological defect points identified from 2D dissections along all possible directions, we computed the local topological charge imbalance $\bar{C}^c_{k}$ (scaled within the range [-1,1]) within these cubical cells. 

\begin{figure}[h]                                   
\centering
\includegraphics[width=1\linewidth]{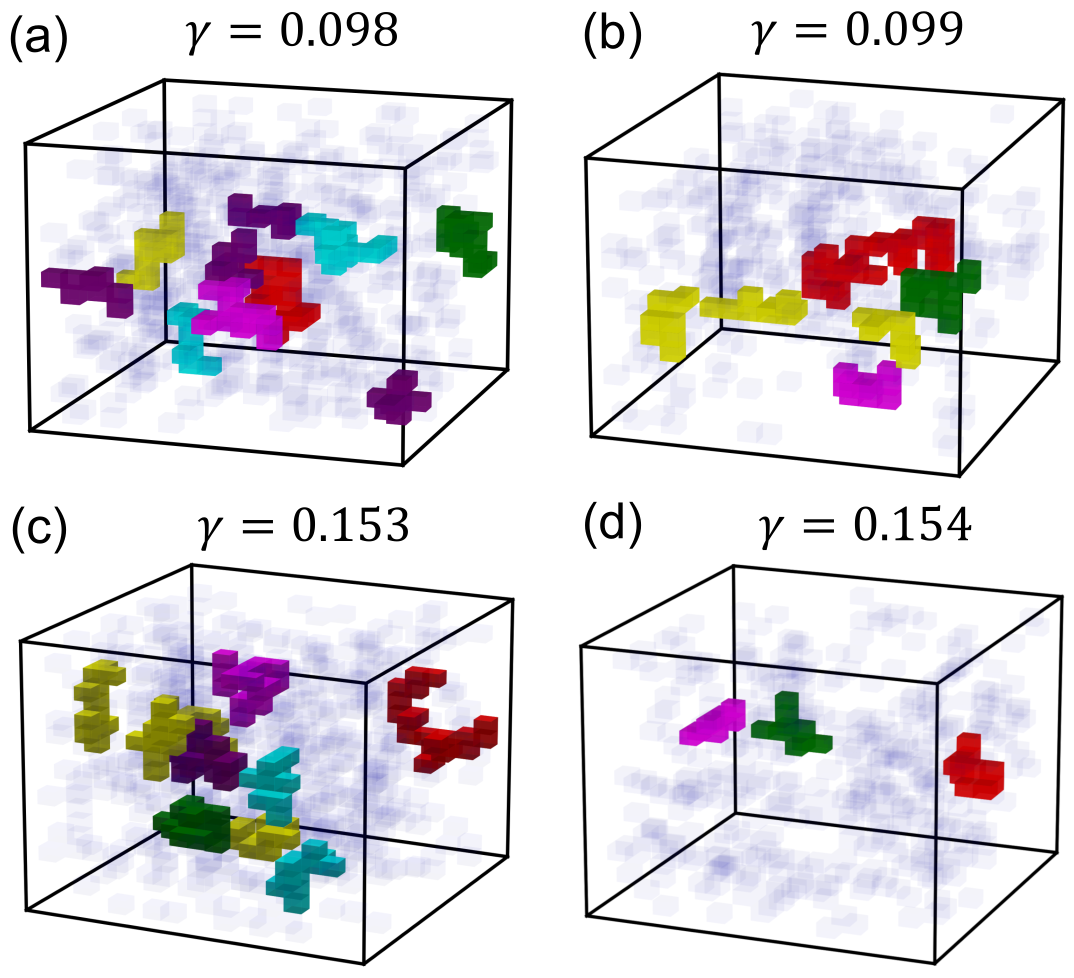}
\caption{The simulation boxes contain cubical cells with negative charge ($\bar{C}^c_{k}<-0.6$) shaded in light blue. Within this representation, we present clusters with size $n_c \ge 8$ from this negative charge cloud. \textbf{(a)} At $\gamma=0.098$ (just before yielding), 11 clusters are displayed, colored in red ($n_c=17$, $N_{n_c}=1$), green ($n_c=15$, $N_{n_c}=1$), magenta ($n_c=14$, $N_{n_c}=1$), yellow ($n_c=10$, $N_{n_c}=1$), cyan ($n_c=9$, $N_{n_c}=3$), and purple ($n_c=8$, $N_{n_c}=4$), respectively. \textbf{(b)} At $\gamma=0.099$ (at yielding), the 6 largest clusters are shown in red ($n_c=31$, $N_{n_c}=1$), green ($n_c=13$, $N_{n_c}=1$), magenta ($n_c=10$, $N_{n_c}=1$), and yellow ($n_c=9$, $N_{n_c}=3$). \textbf{(c)} At $\gamma=0.153$ (just before another PE), 11 largest clusters are presented in red ($n_c=21$, $N_{n_c}=1$), green ($n_c=17$, $N_{n_c}=1$), magenta ($n_c=16$, $N_{n_c}=1$), yellow ($n_c=12$, $N_{n_c}=4$), cyan ($n_c=11$, $N_{n_c}=2$), and purple ($n_c=8$, $N_{n_c}=2$). \textbf{(d)} At $\gamma=0.154$ (occurrence of another PE), 3 clusters are displayed in red ($n_c=10$, $N_{n_c}=1$), green ($n_c=9$, $N_{n_c}=1$), magenta ($n_c=8$, $N_{n_c}=1$), respectively. Note that $N_{n_c}$ is the number of clusters of size $n_c$.}
\label{fig7}
\end{figure}

In Fig.\ref{fig7}, 3D simulation boxes are depicted, wherein coarse-grained cells with $\bar{C}^c_{k}<-0.6$ are highlighted in light blue. Additionally, we identify connected cubical cells with large negative charges, referred to as clusters. In Table \ref{tab1}, we present the number of clusters $N_{cls}$ with a size of $n_c \ge 8$ just before a PE, at the PE, and after the PE for various values of $\gamma$. In all cases, we observe a significant decrease in $N_{\text{cls}}$ at PEs, followed by an increase after each PE. Furthermore, Fig.\ref{fig7} (a)-(b) show all $11$ clusters with $n_c \ge 8$ just before yielding and all $6$ clusters at the PE. A similar scenario is depicted in Fig.\ref{fig7} (c)-(d) for another PE at $\gamma=0.154$. Notably, we observe the disappearance of most of these large connected clusters at PEs and their subsequent development just after the PEs. This suggests that the clustering of negative topological charges is a precursor of a PE and that PEs can be identified with the sudden fragmentation of these large clusters of negative topological charge.  
\begin{table}[h]
\begin{center}
    \caption{The number of clusters $N_{cls}$ having size $n_c\ge 8$ for several plastic events.}
    \small 
    \centering
    \begin{tabular}{|p{4cm}|c|c|c|}
        \hline
        \multicolumn{1}{|c|}{Strain~$\gamma$ at PE} & {$N_{\text{cls}}$ (before)} & {$N_{\text{cls}}$ (at PE)} & {$N_{\text{cls}}$ (after)}\\
        \hline
        \multicolumn{1}{|c|}{$0.099$} & $11$ & 6 & 13\\
        \hline
        \multicolumn{1}{|c|}{$0.106$} & $15$ & 4 & 8\\
        \hline
        \multicolumn{1}{|c|}{$0.120$} & $7$ & 4 & 8\\
        \hline
        \multicolumn{1}{|c|}{$0.126$} & $8$ & 4 & 9\\
        \hline
        \multicolumn{1}{|c|}{$0.154$} & $11$ & 3 & 9\\
         \hline
    \end{tabular}
    \label{tab1}
    \end{center}
\end{table}

\begin{figure}[h]                                     
\centering
\includegraphics[width=\linewidth]{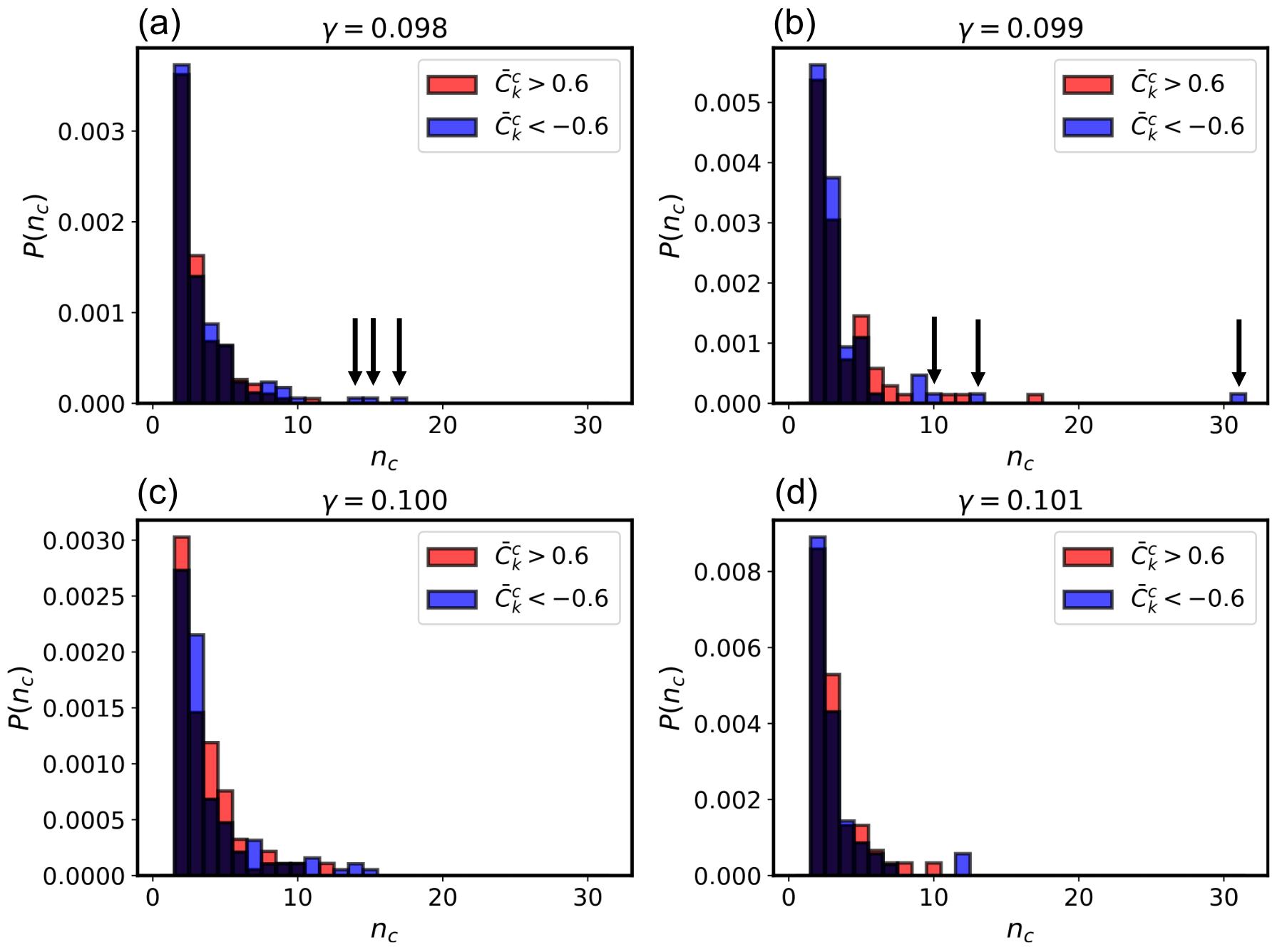}
\caption{\textbf{(a)-(d)} The normalised cluster size distribution $P(n_c)$~($n_c$ being the size of the cluster, \textit{i.e.}, number of connected cubical cells) are plotted for positive and negative charges at different values of strain $\gamma$. Note that the overlapping regions between the distributions for positive and negative charges are plotted in black. The distributions have been normalised in such a way that the area under the curve is unity.}
\label{fig8}
\end{figure}

We examine the spatial distribution and probability distribution of connected cells that had high net charge imbalance of the same sign. We then calculate the normalized cluster size distribution $P(n_c)$, where $n_c$ represents the number of connected cubical cells with a high topological charge of the same sign. In Fig.\ref{fig8}(a)-(d), we depict the cluster size distribution $P(n_c)$ for $\bar{C}^c_k > 0.6$ and $\bar{C}^c_k<-0.6$ during yielding. We have highlighted the three largest clusters in the distributions for $\gamma=0.098$ and $\gamma=0.099$ in Fig. \ref{fig8}(a)-(b). Right before yielding, we observe the emergence of several large clusters, with the largest one being of negative net charge imbalance. At yielding, many of these large clusters disappear and the distribution tends to narrow and fragment into clusters of smaller sizes. After each PE, the system again tends to develop negative charged clusters (see table \ref{tab1}). The rapid formation and disappearance of large clusters of negative topological charge -- a ``topological avalanche'' \cite{https://doi.org/10.1103/RevModPhys.76.471}-- suggests that yielding might be caused by the massive annhilation of topological defects. Furthermore, following the recent results presented in \cite{desmarchelier2024topological}, the large clusters of negative topological charge, related to STZ, appearing before yielding might be a manifestation of the formation of shear bands. It is also plausible that the structure and dynamics of these defects bear strong connections with the concept of elastic screening discussed in \cite{PhysRevE.104.024904}. 
 
\section*{Conclusion}
In this study, we systematically explored the formation, dynamics and statistical properties of topological defects in a 3D polymer glass model under shear deformation. 
We first established that, in 3D glasses, the topological defects with -1 charge (saddles or anti-vortices) are associated with large non-affine displacements and act as triggers of local plastic activity. Also, the total defects density anti-correlates with the non-affine displacements magnitude as a function of shear strain. This suggests that the transient accumulation of defects triggers a subsequent plastic event, upon further increasing the strain, via massive annihilation of negative and positive defects. Overall, the deformation process upon increasing the strain is characterized by a dominance of negative charges over positive charges, which is linked to the tendency of the polymer glass to strain-soften prior and across yielding. Through a coarse-graining procedure into mesoscopic cells, we unveiled the formation of several large clusters bearing net negative topological charge before the yielding point. These clusters are associated with huge values of the non-affine displacement magnitude. The emergence of these large negative clusters before yielding is followed by its massive annihilation right at yielding, which is accompanied by a system-spanning plastic event marking the onset of plastic flow. While our findings pertain to polymer glasses, we expect the same methodology to yield interesting insights into the deformation behaviour of other amorphous and glassy systems.

Several open questions remain to be answered. As already anticipated in \cite{desmarchelier2024topological}, topological defects with negative charge exhibit strong quasi-localized quadrupolar field previously identified as the carriers of plasticity in amorphous solids \cite{Sengupta,Wilde}. Indeed, topological defects can be identified as geometric singularities even in amorphous systems \cite{Moshe,PhysRevE.107.055004}, and might constitute the building blocks for the dipolar and quadrupolar field discussed in the literature in relation to elastic screening and anomalous elasticity \cite{kumar2023elasticity}. These topological defects could provide the missing link to understand yielding in amorphous systems as a topological phase transition \cite{PhysRevE.107.055004,jin2024intermediate}, akin to the famous Berezinskii-Kosterlitz-Thouless (BKT) transition in 2D crystals \cite{PhysRevB.22.2514,Fraggedakis}. In our study, we characterized the topological defects by analyzing quasi-2D slices of the three-dimensional system. It would be interesting to investigate 3D defects within the fully three-dimensional non-affine field. This direct 3D characterization of topological defects in glasses has been less explored in the literature (see nevertheless \cite{Cao2018}) and is computationally more challenging. This presents an intriguing avenue for future research.\\

Vortex-like topological defects have been originally proposed in the eigenvector field \cite{wu2023}, while our study generalizes this concept to the displacement vector field. At this point, it is not entirely clear which are the differences and similarities between the two approaches. Nevertheless, following the aforementioned results of \cite{PhysRevB.66.174205} it is plausible to expect a correlation between the topological features in the displacement field and those in the low-frequency eigenvector fields around strong plastic events. The main reason behind this expected correlation is the fact that plastic events are typically dominated by low-frequency modes that correspond to the disappearance of local potential barriers in the potential energy landscape. A more detailed comparison between these two methods is left for future research. See also \cite{desmarchelier2024topological} for an analysis of topological defects in the displacement vector field.\\

Moreover, topological defects may provide a geometric definition for the liquid-like atomic environments \cite{PhysRevB.72.245205} and display a strong connection with the emergence of unstable instantaneous modes, already associated in the literature to plastic instabilities \cite{PhysRevE.105.055004,PhysRevLett.127.108003}. On the other hand, it would be interesting to analyze from a theoretical point of view how these topological defects manifest themselves within the tensor gauge theory for anomalous elasticity of \cite{PhysRevLett.125.118002,PhysRevE.106.065004}.\\

Finally, to give more credibility to this novel approach, it is imperative to identify these topological defects experimentally. Colloidal systems or granular matter systems seem to be the most promising playground for this scope. Microbeam X-ray scattering techniques, as used in \cite{doi:10.1126/sciadv.abn0681}, might also be able detect the signatures of these topological defects. Finally, X-ray irradiation \cite{PhysRevX.13.041031} could provide a method to induce and control the dynamics of the latter.\\

As a final remark, despite what has been thought for long time, topological concepts are still valid beyond the realm of structural order. Here we show topological defects emerge as extremely promising candidate to characterize plasticity and mechanical failure in amorphous materials at a microscopic level, as achieved for their crystalline counterparts. 

\section*{Appendix: Simulation methods}
We use the Kremer-Grest model \cite{grest1986} to investigate a coarse-grained polymer system consisting of linear chains, each composed of $50$ monomers. The monomer masses within a polymer chain alternate between two values, $m_1 = 1$ and $m_2 = 3$, with a total of $N = 10,000$ monomers in the system.

In our system, monomers interact through the potential:
\begin{equation}\label{full_lj}
U_{\text{LJ}}(r)=
\begin{cases}
u(r)-u(r_c), & \text{for}~r<r_c=2.5\sigma,\\
0, & \text{otherwise.}
\end{cases}
\end{equation}
Here, $u(r)$ represents the standard Lennard-Jones (LJ) potential
\begin{equation}\label{ljII}
u(r) = 4{\varepsilon} \left [ \left (\frac{\sigma}{r} \right )^{12} - \left (\frac{\sigma}{r} \right )^6 \right ],
\end{equation}
where $\varepsilon$ is the interaction strength, and $\sigma$ is the diameter of each monomer. The inter-particle distance $r$ is constrained to be within the interaction range $r_c = 2.5\sigma$. Additionally, the covalent along-the-chain bonds are represented by a finite extensible nonlinear elastic (FENE) potential \cite{grest1986}:
\begin{equation}\label{ljI}
U_{\text{FENE}}(r) = -\frac{Kr_0^2}{2} \ln \left [ 1 - \left (\frac{r}{r_0} \right )^2\right ].
\end{equation}
Here, we set the strength of the FENE interaction as $K = 30$ and the range $r_0 = 1.5\sigma$.

The polymer chains were placed within a three-dimensional cubic box and subjected to periodic boundary conditions in all directions. Equilibration of the system was conducted using the LAMMPS simulation package \cite{lammps}. For the deformation process, we employed an athermal quasi-static (AQS) protocol utilizing the LAMMPS \cite{lammps} environment. The glass sample is prepared at zero temperature and subsequently subjected to a quasi-static shear procedure with each strain step of $\delta \gamma_{xz} = 0.001$. We expressed mass, length, and time in units of $m_1$, $\sigma$, and $\tau = ({m_1\sigma^2/{\varepsilon}})^{1/2}$. In our study, we set $m_1$, $\sigma$, and $\varepsilon$ to unity. In all our simulations the chains are treated as fully-flexible, with no covalent-like bond-bending terms in the potential.

We systematically partitioned the simulation box into $20\sigma \times 20\sigma$ 2D slices of width $2\sigma$ along the $xz$-plane into a $80 \times 80$ square lattice. Subsequently, we executed an interpolation of non-affine displacement values at these lattice grid points. The interpolation has been performed using an in-built Python function ``griddata'' that takes as input the displacement field scattered data and gives the displacement field at each constructed lattice grid. At each grid point, the phase-angle $\theta$ of the non-affine field is determined. Utilizing nearest neighbor grid points, we systematically constructed closed square loops and compute the topological charge $q$ at the central point of each such square loop using \eqref{topo_charge}. 

When approximating the contour integration with a discrete Riemann sum, precise numerical evaluation requires numerous points along the path or, equivalently, a small step between summation points. To isolate single defects in the plane, the circulation path can be contracted to the immediate vicinity, maintaining the topological index of the (potential) vortex core. Ensuring small steps between points is achieved by interpolating the smooth and continuous field near the suspected vortex core, preserving the topology. For efficient upscaling through interpolation, only a few points need sampling, as they are in close proximity. On a Cartesian grid, four points over a $2 \times 2$ pixel grid are numerically convenient. In clustered vortex scenarios, expanded circulation paths might inaccurately identify unresolved higher-order vortices (by encompassing multiple like-signed topological charges) or overlook vortices altogether (in the presence of an equal number of $-ve$ and $+ve$ signed vortices within the contour).

\subsection*{Supplementary Material}
Supplementary material is available at PNAS Nexus online.
\subsection*{Data availability}
All data for this study are presented in the manuscript and the Supplementary material. Additional data (polymer glass models and interpolated fields) are available from the Figshare repository: DOI: 10.26180/26199917.
\subsection*{Funding}
M.B. acknowledge the support of the Shanghai Municipal Science and Technology Major Project (Grant No.2019SHZDZX01). MB acknowledges the support of the sponsorship from the Yangyang Development Fund. ACYL acknowledges support from the Australian Research Council (FT180100594). A.Z. gratefully acknowledges funding from the European Union through Horizon Europe ERC Grant number: 101043968 ``Multimech'', and from the Nieders{\"a}chsische Akademie der Wissenschaften zu G{\"o}ttingen in the frame of the Gauss Professorship program. A.Z. and A.B. gratefully acknowledge funding from US Army Research Office through contract nr. W911NF-22-2-0256.

\subsection*{Author contributions statement}
A.~Bera performed the numerical simulations and data analysis with the help of A.~Liu. A.~Bera wrote the manuscript with the help of the other authors. A.~Zaccone supervised the project. All authors contributed to the development of the ideas behind this work and the physical interpretation of the results. 

\acknowledgements{We would like to thank Y.~Jin, D.~Pan, W.~Zhang, J.~Zhang, M.~Pica Ciamarra, M.~Moshe and C.~Jiang for useful discussions and collaboration on related topics.}


\bibliography{bib}

\begin{thebibliography}{50}
\expandafter\ifx\csname natexlab\endcsname\relax\def\natexlab#1{#1}\fi
\expandafter\ifx\csname bibnamefont\endcsname\relax
  \def\bibnamefont#1{#1}\fi
\expandafter\ifx\csname bibfnamefont\endcsname\relax
  \def\bibfnamefont#1{#1}\fi
\expandafter\ifx\csname citenamefont\endcsname\relax
  \def\citenamefont#1{#1}\fi
\expandafter\ifx\csname url\endcsname\relax
  \def\url#1{\texttt{#1}}\fi
\expandafter\ifx\csname urlprefix\endcsname\relax\def\urlprefix{URL }\fi
\providecommand{\bibinfo}[2]{#2}
\providecommand{\eprint}[2][]{\url{#2}}

\bibitem[{\citenamefont{Taylor}(1934)}]{Taylor}
\bibinfo{author}{\bibfnamefont{G.~I.} \bibnamefont{Taylor}}, \bibinfo{journal}{Proceedings of the Royal Society of London. Series A, Containing Papers of a Mathematical and Physical Character} \textbf{\bibinfo{volume}{145}}, \bibinfo{pages}{362} (\bibinfo{year}{1934}).

\bibitem[{\citenamefont{Wang et~al.}(2015)\citenamefont{Wang, Yang, Nieh, and Liu}}]{WANG201581}
\bibinfo{author}{\bibfnamefont{W.~H.} \bibnamefont{Wang}}, \bibinfo{author}{\bibfnamefont{Y.}~\bibnamefont{Yang}}, \bibinfo{author}{\bibfnamefont{T.~G.} \bibnamefont{Nieh}}, \bibnamefont{and} \bibinfo{author}{\bibfnamefont{C.~T.} \bibnamefont{Liu}}, \bibinfo{journal}{Intermetallics} \textbf{\bibinfo{volume}{67}}, \bibinfo{pages}{81} (\bibinfo{year}{2015}).

\bibitem[{\citenamefont{Gilman}(2003)}]{10.1063/1.1662243}
\bibinfo{author}{\bibfnamefont{J.~J.} \bibnamefont{Gilman}}, \bibinfo{journal}{Journal of Applied Physics} \textbf{\bibinfo{volume}{44}}, \bibinfo{pages}{675} (\bibinfo{year}{2003}).

\bibitem[{\citenamefont{Chaudhari et~al.}(1979)\citenamefont{Chaudhari, Levi, and Steinhardt}}]{PhysRevLett.43.1517}
\bibinfo{author}{\bibfnamefont{P.}~\bibnamefont{Chaudhari}}, \bibinfo{author}{\bibfnamefont{A.}~\bibnamefont{Levi}}, \bibnamefont{and} \bibinfo{author}{\bibfnamefont{P.}~\bibnamefont{Steinhardt}}, \bibinfo{journal}{Phys. Rev. Lett.} \textbf{\bibinfo{volume}{43}}, \bibinfo{pages}{1517} (\bibinfo{year}{1979}).

\bibitem[{\citenamefont{Steinhardt and Chaudhari}(1981)}]{doi:10.1080/01418618108235816}
\bibinfo{author}{\bibfnamefont{P.~J.} \bibnamefont{Steinhardt}} \bibnamefont{and} \bibinfo{author}{\bibfnamefont{P.}~\bibnamefont{Chaudhari}}, \bibinfo{journal}{Philosophical Magazine A} \textbf{\bibinfo{volume}{44}}, \bibinfo{pages}{1375} (\bibinfo{year}{1981}).

\bibitem[{\citenamefont{Egami et~al.}(1980)\citenamefont{Egami, Maeda, and Vitek}}]{doi:10.1080/01418618008243894}
\bibinfo{author}{\bibfnamefont{T.}~\bibnamefont{Egami}}, \bibinfo{author}{\bibfnamefont{K.}~\bibnamefont{Maeda}}, \bibnamefont{and} \bibinfo{author}{\bibfnamefont{V.}~\bibnamefont{Vitek}}, \bibinfo{journal}{Philosophical Magazine A} \textbf{\bibinfo{volume}{41}}, \bibinfo{pages}{883} (\bibinfo{year}{1980}).

\bibitem[{\citenamefont{Alexander}(1998)}]{ALEXANDER199865}
\bibinfo{author}{\bibfnamefont{S.}~\bibnamefont{Alexander}}, \bibinfo{journal}{Physics Reports} \textbf{\bibinfo{volume}{296}}, \bibinfo{pages}{65} (\bibinfo{year}{1998}).

\bibitem[{\citenamefont{Falk and Langer}(1998)}]{PhysRevE.57.7192}
\bibinfo{author}{\bibfnamefont{M.~L.} \bibnamefont{Falk}} \bibnamefont{and} \bibinfo{author}{\bibfnamefont{J.~S.} \bibnamefont{Langer}}, \bibinfo{journal}{Phys. Rev. E} \textbf{\bibinfo{volume}{57}}, \bibinfo{pages}{7192} (\bibinfo{year}{1998}).

\bibitem[{\citenamefont{Fraggedakis et~al.}(2023)\citenamefont{Fraggedakis, Hasyim, and Mandadapu}}]{Fraggedakis}
\bibinfo{author}{\bibfnamefont{D.}~\bibnamefont{Fraggedakis}}, \bibinfo{author}{\bibfnamefont{M.~R.} \bibnamefont{Hasyim}}, \bibnamefont{and} \bibinfo{author}{\bibfnamefont{K.~K.} \bibnamefont{Mandadapu}}, \bibinfo{journal}{Proc. Natl. Acad. Sci. U.S.A.} \textbf{\bibinfo{volume}{120}}, \bibinfo{pages}{e2209144120} (\bibinfo{year}{2023}).

\bibitem[{\citenamefont{Nicolas et~al.}(2018)\citenamefont{Nicolas, Ferrero, Martens, and Barrat}}]{RevModPhys.90.045006}
\bibinfo{author}{\bibfnamefont{A.}~\bibnamefont{Nicolas}}, \bibinfo{author}{\bibfnamefont{E.~E.} \bibnamefont{Ferrero}}, \bibinfo{author}{\bibfnamefont{K.}~\bibnamefont{Martens}}, \bibnamefont{and} \bibinfo{author}{\bibfnamefont{J.-L.} \bibnamefont{Barrat}}, \bibinfo{journal}{Rev. Mod. Phys.} \textbf{\bibinfo{volume}{90}}, \bibinfo{pages}{045006} (\bibinfo{year}{2018}).

\bibitem[{\citenamefont{Manning and Liu}(2011)}]{PhysRevLett.107.108302}
\bibinfo{author}{\bibfnamefont{M.~L.} \bibnamefont{Manning}} \bibnamefont{and} \bibinfo{author}{\bibfnamefont{A.~J.} \bibnamefont{Liu}}, \bibinfo{journal}{Phys. Rev. Lett.} \textbf{\bibinfo{volume}{107}}, \bibinfo{pages}{108302} (\bibinfo{year}{2011}).

\bibitem[{\citenamefont{Fumeron and Berche}(2023)}]{Fumeron2023}
\bibinfo{author}{\bibfnamefont{S.}~\bibnamefont{Fumeron}} \bibnamefont{and} \bibinfo{author}{\bibfnamefont{B.}~\bibnamefont{Berche}}, \bibinfo{journal}{Eur. Phys. J. Spec. Top.} \textbf{\bibinfo{volume}{232}}, \bibinfo{pages}{1813} (\bibinfo{year}{2023}).

\bibitem[{\citenamefont{Lerner and Bouchbinder}(2021)}]{10.1063/5.0069477}
\bibinfo{author}{\bibfnamefont{E.}~\bibnamefont{Lerner}} \bibnamefont{and} \bibinfo{author}{\bibfnamefont{E.}~\bibnamefont{Bouchbinder}}, \bibinfo{journal}{J. Chem. Phys.} \textbf{\bibinfo{volume}{155}}, \bibinfo{pages}{200901} (\bibinfo{year}{2021}).

\bibitem[{\citenamefont{Lema\^{\i}tre et~al.}(2021)\citenamefont{Lema\^{\i}tre, Mondal, Moshe, Procaccia, Roy, and Screiber-Re'em}}]{PhysRevE.104.024904}
\bibinfo{author}{\bibfnamefont{A.}~\bibnamefont{Lema\^{\i}tre}}, \bibinfo{author}{\bibfnamefont{C.}~\bibnamefont{Mondal}}, \bibinfo{author}{\bibfnamefont{M.}~\bibnamefont{Moshe}}, \bibinfo{author}{\bibfnamefont{I.}~\bibnamefont{Procaccia}}, \bibinfo{author}{\bibfnamefont{S.}~\bibnamefont{Roy}}, \bibnamefont{and} \bibinfo{author}{\bibfnamefont{K.}~\bibnamefont{Screiber-Re'em}}, \bibinfo{journal}{Phys. Rev. E} \textbf{\bibinfo{volume}{104}}, \bibinfo{pages}{024904} (\bibinfo{year}{2021}).

\bibitem[{\citenamefont{Lund}(2015)}]{PhysRevB.91.094102}
\bibinfo{author}{\bibfnamefont{F.}~\bibnamefont{Lund}}, \bibinfo{journal}{Phys. Rev. B} \textbf{\bibinfo{volume}{91}}, \bibinfo{pages}{094102} (\bibinfo{year}{2015}).

\bibitem[{\citenamefont{Zhou et~al.}(2023)\citenamefont{Zhou, Sun, Gao, Wang, and Yu}}]{ZHOU2023118701}
\bibinfo{author}{\bibfnamefont{Z.-Y.} \bibnamefont{Zhou}}, \bibinfo{author}{\bibfnamefont{Y.}~\bibnamefont{Sun}}, \bibinfo{author}{\bibfnamefont{L.}~\bibnamefont{Gao}}, \bibinfo{author}{\bibfnamefont{Y.-J.} \bibnamefont{Wang}}, \bibnamefont{and} \bibinfo{author}{\bibfnamefont{H.-B.} \bibnamefont{Yu}}, \bibinfo{journal}{Acta Materialia} \textbf{\bibinfo{volume}{246}}, \bibinfo{pages}{118701} (\bibinfo{year}{2023}).

\bibitem[{\citenamefont{Richard et~al.}(2020)\citenamefont{Richard, Ozawa, Patinet, Stanifer, Shang, Ridout, Xu, Zhang, Morse, Barrat et~al.}}]{PhysRevMaterials.4.113609}
\bibinfo{author}{\bibfnamefont{D.}~\bibnamefont{Richard}}, \bibinfo{author}{\bibfnamefont{M.}~\bibnamefont{Ozawa}}, \bibinfo{author}{\bibfnamefont{S.}~\bibnamefont{Patinet}}, \bibinfo{author}{\bibfnamefont{E.}~\bibnamefont{Stanifer}}, \bibinfo{author}{\bibfnamefont{B.}~\bibnamefont{Shang}}, \bibinfo{author}{\bibfnamefont{S.~A.} \bibnamefont{Ridout}}, \bibinfo{author}{\bibfnamefont{B.}~\bibnamefont{Xu}}, \bibinfo{author}{\bibfnamefont{G.}~\bibnamefont{Zhang}}, \bibinfo{author}{\bibfnamefont{P.~K.} \bibnamefont{Morse}}, \bibinfo{author}{\bibfnamefont{J.-L.} \bibnamefont{Barrat}}, \bibnamefont{et~al.}, \bibinfo{journal}{Phys. Rev. Mater.} \textbf{\bibinfo{volume}{4}}, \bibinfo{pages}{113609} (\bibinfo{year}{2020}).

\bibitem[{\citenamefont{Baggioli et~al.}(2021)\citenamefont{Baggioli, Kriuchevskyi, Sirk, and Zaccone}}]{baggioli2021}
\bibinfo{author}{\bibfnamefont{M.}~\bibnamefont{Baggioli}}, \bibinfo{author}{\bibfnamefont{I.}~\bibnamefont{Kriuchevskyi}}, \bibinfo{author}{\bibfnamefont{T.~W.} \bibnamefont{Sirk}}, \bibnamefont{and} \bibinfo{author}{\bibfnamefont{A.}~\bibnamefont{Zaccone}}, \bibinfo{journal}{Phys. Rev. Lett.} \textbf{\bibinfo{volume}{127}}, \bibinfo{pages}{015501} (\bibinfo{year}{2021}).

\bibitem[{\citenamefont{Tanguy et~al.}(2002)\citenamefont{Tanguy, Wittmer, Leonforte, and Barrat}}]{PhysRevB.66.174205}
\bibinfo{author}{\bibfnamefont{A.}~\bibnamefont{Tanguy}}, \bibinfo{author}{\bibfnamefont{J.~P.} \bibnamefont{Wittmer}}, \bibinfo{author}{\bibfnamefont{F.}~\bibnamefont{Leonforte}}, \bibnamefont{and} \bibinfo{author}{\bibfnamefont{J.-L.} \bibnamefont{Barrat}}, \bibinfo{journal}{Phys. Rev. B} \textbf{\bibinfo{volume}{66}}, \bibinfo{pages}{174205} (\bibinfo{year}{2002}).

\bibitem[{\citenamefont{DiDonna and Lubensky}(2005)}]{PhysRevE.72.066619}
\bibinfo{author}{\bibfnamefont{B.~A.} \bibnamefont{DiDonna}} \bibnamefont{and} \bibinfo{author}{\bibfnamefont{T.~C.} \bibnamefont{Lubensky}}, \bibinfo{journal}{Phys. Rev. E} \textbf{\bibinfo{volume}{72}}, \bibinfo{pages}{066619} (\bibinfo{year}{2005}).

\bibitem[{\citenamefont{Zaccone}(2023)}]{zaccone2023}
\bibinfo{author}{\bibfnamefont{A.}~\bibnamefont{Zaccone}}, \emph{\bibinfo{title}{Theory of Disordered Solids}} (\bibinfo{publisher}{Springer}, \bibinfo{address}{Heidelberg}, \bibinfo{year}{2023}).

\bibitem[{\citenamefont{Kleinert}(1989)}]{kleinert1989gauge}
\bibinfo{author}{\bibfnamefont{H.}~\bibnamefont{Kleinert}}, \emph{\bibinfo{title}{Gauge Fields in Condensed Matter}}, \bibinfo{number}{v. 2} (\bibinfo{publisher}{World Scientific}, \bibinfo{year}{1989}).

\bibitem[{\citenamefont{Baggioli et~al.}(2022)\citenamefont{Baggioli, Landry, and Zaccone}}]{baggioli2022}
\bibinfo{author}{\bibfnamefont{M.}~\bibnamefont{Baggioli}}, \bibinfo{author}{\bibfnamefont{M.}~\bibnamefont{Landry}}, \bibnamefont{and} \bibinfo{author}{\bibfnamefont{A.}~\bibnamefont{Zaccone}}, \bibinfo{journal}{Phys. Rev. E} \textbf{\bibinfo{volume}{105}}, \bibinfo{pages}{024602} (\bibinfo{year}{2022}).

\bibitem[{\citenamefont{Wu et~al.}(2023)\citenamefont{Wu, Chen, Wang, Kob, and Xu}}]{wu2023}
\bibinfo{author}{\bibfnamefont{Z.~W.} \bibnamefont{Wu}}, \bibinfo{author}{\bibfnamefont{Y.}~\bibnamefont{Chen}}, \bibinfo{author}{\bibfnamefont{W.-H.} \bibnamefont{Wang}}, \bibinfo{author}{\bibfnamefont{W.}~\bibnamefont{Kob}}, \bibnamefont{and} \bibinfo{author}{\bibfnamefont{L.}~\bibnamefont{Xu}}, \bibinfo{journal}{Nature Communications} \textbf{\bibinfo{volume}{14}}, \bibinfo{pages}{2955} (\bibinfo{year}{2023}).

\bibitem[{\citenamefont{Baggioli}(2023)}]{Baggioli2023}
\bibinfo{author}{\bibfnamefont{M.}~\bibnamefont{Baggioli}}, \bibinfo{journal}{Nat. Commun.} \textbf{\bibinfo{volume}{14}}, \bibinfo{pages}{2956} (\bibinfo{year}{2023}).

\bibitem[{\citenamefont{\ifmmode~\mbox{\c{S}}\else \c{S}\fi{}opu et~al.}(2017)\citenamefont{\ifmmode~\mbox{\c{S}}\else \c{S}\fi{}opu, Stukowski, Stoica, and Scudino}}]{PhysRevLett.119.195503}
\bibinfo{author}{\bibfnamefont{D.}~\bibnamefont{\ifmmode~\mbox{\c{S}}\else \c{S}\fi{}opu}}, \bibinfo{author}{\bibfnamefont{A.}~\bibnamefont{Stukowski}}, \bibinfo{author}{\bibfnamefont{M.}~\bibnamefont{Stoica}}, \bibnamefont{and} \bibinfo{author}{\bibfnamefont{S.}~\bibnamefont{Scudino}}, \bibinfo{journal}{Phys. Rev. Lett.} \textbf{\bibinfo{volume}{119}}, \bibinfo{pages}{195503} (\bibinfo{year}{2017}).

\bibitem[{\citenamefont{Sopu}(2023)}]{SOPU2023170585}
\bibinfo{author}{\bibfnamefont{D.}~\bibnamefont{Sopu}}, \bibinfo{journal}{Journal of Alloys and Compounds} \textbf{\bibinfo{volume}{960}}, \bibinfo{pages}{170585} (\bibinfo{year}{2023}).

\bibitem[{\citenamefont{Desmarchelier et~al.}(2024)\citenamefont{Desmarchelier, Fajardo, and Falk}}]{desmarchelier2024topological}
\bibinfo{author}{\bibfnamefont{P.}~\bibnamefont{Desmarchelier}}, \bibinfo{author}{\bibfnamefont{S.}~\bibnamefont{Fajardo}}, \bibnamefont{and} \bibinfo{author}{\bibfnamefont{M.~L.} \bibnamefont{Falk}}, \bibinfo{journal}{Phys. Rev. Lett.} \textbf{\bibinfo{volume}{109}}, \bibinfo{pages}{L053002} (\bibinfo{year}{2024}).

\bibitem[{\citenamefont{Kleman and Friedel}(2008)}]{RevModPhys.80.61}
\bibinfo{author}{\bibfnamefont{M.}~\bibnamefont{Kleman}} \bibnamefont{and} \bibinfo{author}{\bibfnamefont{J.}~\bibnamefont{Friedel}}, \bibinfo{journal}{Rev. Mod. Phys.} \textbf{\bibinfo{volume}{80}}, \bibinfo{pages}{61} (\bibinfo{year}{2008}).

\bibitem[{\citenamefont{Kupferman et~al.}(2018)\citenamefont{Kupferman, , Maor, and Rosenthal}}]{kupferman2018}
\bibinfo{author}{\bibfnamefont{R.}~\bibnamefont{Kupferman}}, , \bibinfo{author}{\bibfnamefont{C.}~\bibnamefont{Maor}}, \bibnamefont{and} \bibinfo{author}{\bibfnamefont{R.}~\bibnamefont{Rosenthal}}, \bibinfo{journal}{Israel Journal of Mathematics} \textbf{\bibinfo{volume}{223}}, \bibinfo{pages}{75} (\bibinfo{year}{2018}).

\bibitem[{\citenamefont{Yavari and Goriely}(2012)}]{yavari_new}
\bibinfo{author}{\bibfnamefont{A.}~\bibnamefont{Yavari}} \bibnamefont{and} \bibinfo{author}{\bibfnamefont{A.}~\bibnamefont{Goriely}}, \bibinfo{journal}{Proc. R. Soc. A} \textbf{\bibinfo{volume}{223}}, \bibinfo{pages}{3902} (\bibinfo{year}{2012}).

\bibitem[{\citenamefont{Altshuler and Johansen}(2004)}]{https://doi.org/10.1103/RevModPhys.76.471}
\bibinfo{author}{\bibfnamefont{E.}~\bibnamefont{Altshuler}} \bibnamefont{and} \bibinfo{author}{\bibfnamefont{T.~H.} \bibnamefont{Johansen}}, \bibinfo{journal}{Reviews of Modern Physics} \textbf{\bibinfo{volume}{76}}, \bibinfo{pages}{471} (\bibinfo{year}{2004}).

\bibitem[{\citenamefont{Kremer and Grest}(1986)}]{grest1986}
\bibinfo{author}{\bibfnamefont{K.}~\bibnamefont{Kremer}} \bibnamefont{and} \bibinfo{author}{\bibfnamefont{G.~S.} \bibnamefont{Grest}}, \bibinfo{journal}{Phys. Rev. A} \textbf{\bibinfo{volume}{33}}, \bibinfo{pages}{3628} (\bibinfo{year}{1986}).

\bibitem[{\citenamefont{Pearson and Galton}(1895)}]{pearson1895}
\bibinfo{author}{\bibfnamefont{K.}~\bibnamefont{Pearson}} \bibnamefont{and} \bibinfo{author}{\bibfnamefont{F.}~\bibnamefont{Galton}}, \bibinfo{journal}{Proceedings of the Royal Society of London} \textbf{\bibinfo{volume}{58}}, \bibinfo{pages}{240} (\bibinfo{year}{1895}).

\bibitem[{\citenamefont{Dasgupta et~al.}(2012)\citenamefont{Dasgupta, Hentschel, and Procaccia}}]{Sengupta}
\bibinfo{author}{\bibfnamefont{R.}~\bibnamefont{Dasgupta}}, \bibinfo{author}{\bibfnamefont{H.~G.~E.} \bibnamefont{Hentschel}}, \bibnamefont{and} \bibinfo{author}{\bibfnamefont{I.}~\bibnamefont{Procaccia}}, \bibinfo{journal}{Phys. Rev. Lett.} \textbf{\bibinfo{volume}{109}}, \bibinfo{pages}{255502} (\bibinfo{year}{2012}).

\bibitem[{\citenamefont{Hieronymus-Schmidt et~al.}(2017)\citenamefont{Hieronymus-Schmidt, R\"osner, Wilde, and Zaccone}}]{Wilde}
\bibinfo{author}{\bibfnamefont{V.}~\bibnamefont{Hieronymus-Schmidt}}, \bibinfo{author}{\bibfnamefont{H.}~\bibnamefont{R\"osner}}, \bibinfo{author}{\bibfnamefont{G.}~\bibnamefont{Wilde}}, \bibnamefont{and} \bibinfo{author}{\bibfnamefont{A.}~\bibnamefont{Zaccone}}, \bibinfo{journal}{Phys. Rev. B} \textbf{\bibinfo{volume}{95}}, \bibinfo{pages}{134111} (\bibinfo{year}{2017}).

\bibitem[{\citenamefont{Moshe et~al.}(2015)\citenamefont{Moshe, Levin, Aharoni, Kupferman, and Sharon}}]{Moshe}
\bibinfo{author}{\bibfnamefont{M.}~\bibnamefont{Moshe}}, \bibinfo{author}{\bibfnamefont{I.}~\bibnamefont{Levin}}, \bibinfo{author}{\bibfnamefont{H.}~\bibnamefont{Aharoni}}, \bibinfo{author}{\bibfnamefont{R.}~\bibnamefont{Kupferman}}, \bibnamefont{and} \bibinfo{author}{\bibfnamefont{E.}~\bibnamefont{Sharon}}, \bibinfo{journal}{Proc. Natl. Acad. Sci. U.S.A.} \textbf{\bibinfo{volume}{112}}, \bibinfo{pages}{10873} (\bibinfo{year}{2015}).

\bibitem[{\citenamefont{Livne et~al.}(2023)\citenamefont{Livne, Schiller, and Moshe}}]{PhysRevE.107.055004}
\bibinfo{author}{\bibfnamefont{N.~S.} \bibnamefont{Livne}}, \bibinfo{author}{\bibfnamefont{A.}~\bibnamefont{Schiller}}, \bibnamefont{and} \bibinfo{author}{\bibfnamefont{M.}~\bibnamefont{Moshe}}, \bibinfo{journal}{Phys. Rev. E} \textbf{\bibinfo{volume}{107}}, \bibinfo{pages}{055004} (\bibinfo{year}{2023}).

\bibitem[{\citenamefont{Kumar and Procaccia}(2023)}]{kumar2023elasticity}
\bibinfo{author}{\bibfnamefont{A.}~\bibnamefont{Kumar}} \bibnamefont{and} \bibinfo{author}{\bibfnamefont{I.}~\bibnamefont{Procaccia}}, \bibinfo{journal}{Europhys. Lett.} \textbf{\bibinfo{volume}{145}}, \bibinfo{pages}{26002} (\bibinfo{year}{2023}).

\bibitem[{\citenamefont{Jin et~al.}(2024)\citenamefont{Jin, Procaccia, and Samanta}}]{jin2024intermediate}
\bibinfo{author}{\bibfnamefont{Y.}~\bibnamefont{Jin}}, \bibinfo{author}{\bibfnamefont{I.}~\bibnamefont{Procaccia}}, \bibnamefont{and} \bibinfo{author}{\bibfnamefont{T.}~\bibnamefont{Samanta}}, \bibinfo{journal}{Phys. Rev. E} \textbf{\bibinfo{volume}{109}}, \bibinfo{pages}{014902} (\bibinfo{year}{2024}).

\bibitem[{\citenamefont{Zippelius et~al.}(1980)\citenamefont{Zippelius, Halperin, and Nelson}}]{PhysRevB.22.2514}
\bibinfo{author}{\bibfnamefont{A.}~\bibnamefont{Zippelius}}, \bibinfo{author}{\bibfnamefont{B.~I.} \bibnamefont{Halperin}}, \bibnamefont{and} \bibinfo{author}{\bibfnamefont{D.~R.} \bibnamefont{Nelson}}, \bibinfo{journal}{Phys. Rev. B} \textbf{\bibinfo{volume}{22}}, \bibinfo{pages}{2514} (\bibinfo{year}{1980}).

\bibitem[{\citenamefont{Cao et~al.}(2018)\citenamefont{Cao, Li, Kou, Xia, Li, Chen, Xie, Xiao, Kob, Hong et~al.}}]{Cao2018}
\bibinfo{author}{\bibfnamefont{Y.}~\bibnamefont{Cao}}, \bibinfo{author}{\bibfnamefont{J.}~\bibnamefont{Li}}, \bibinfo{author}{\bibfnamefont{B.}~\bibnamefont{Kou}}, \bibinfo{author}{\bibfnamefont{C.}~\bibnamefont{Xia}}, \bibinfo{author}{\bibfnamefont{Z.}~\bibnamefont{Li}}, \bibinfo{author}{\bibfnamefont{R.}~\bibnamefont{Chen}}, \bibinfo{author}{\bibfnamefont{H.}~\bibnamefont{Xie}}, \bibinfo{author}{\bibfnamefont{T.}~\bibnamefont{Xiao}}, \bibinfo{author}{\bibfnamefont{W.}~\bibnamefont{Kob}}, \bibinfo{author}{\bibfnamefont{L.}~\bibnamefont{Hong}}, \bibnamefont{et~al.}, \bibinfo{journal}{Nat. Commun.} \textbf{\bibinfo{volume}{9}}, \bibinfo{pages}{2911} (\bibinfo{year}{2018}).

\bibitem[{\citenamefont{Demkowicz and Argon}(2005)}]{PhysRevB.72.245205}
\bibinfo{author}{\bibfnamefont{M.~J.} \bibnamefont{Demkowicz}} \bibnamefont{and} \bibinfo{author}{\bibfnamefont{A.~S.} \bibnamefont{Argon}}, \bibinfo{journal}{Phys. Rev. B} \textbf{\bibinfo{volume}{72}}, \bibinfo{pages}{245205} (\bibinfo{year}{2005}).

\bibitem[{\citenamefont{Kriuchevskyi et~al.}(2022)\citenamefont{Kriuchevskyi, Sirk, and Zaccone}}]{PhysRevE.105.055004}
\bibinfo{author}{\bibfnamefont{I.}~\bibnamefont{Kriuchevskyi}}, \bibinfo{author}{\bibfnamefont{T.~W.} \bibnamefont{Sirk}}, \bibnamefont{and} \bibinfo{author}{\bibfnamefont{A.}~\bibnamefont{Zaccone}}, \bibinfo{journal}{Phys. Rev. E} \textbf{\bibinfo{volume}{105}}, \bibinfo{pages}{055004} (\bibinfo{year}{2022}).

\bibitem[{\citenamefont{Oyama et~al.}(2021)\citenamefont{Oyama, Mizuno, and Ikeda}}]{PhysRevLett.127.108003}
\bibinfo{author}{\bibfnamefont{N.}~\bibnamefont{Oyama}}, \bibinfo{author}{\bibfnamefont{H.}~\bibnamefont{Mizuno}}, \bibnamefont{and} \bibinfo{author}{\bibfnamefont{A.}~\bibnamefont{Ikeda}}, \bibinfo{journal}{Phys. Rev. Lett.} \textbf{\bibinfo{volume}{127}}, \bibinfo{pages}{108003} (\bibinfo{year}{2021}).

\bibitem[{\citenamefont{Nampoothiri et~al.}(2020)\citenamefont{Nampoothiri, Wang, Ramola, Zhang, Bhattacharjee, and Chakraborty}}]{PhysRevLett.125.118002}
\bibinfo{author}{\bibfnamefont{J.~N.} \bibnamefont{Nampoothiri}}, \bibinfo{author}{\bibfnamefont{Y.}~\bibnamefont{Wang}}, \bibinfo{author}{\bibfnamefont{K.}~\bibnamefont{Ramola}}, \bibinfo{author}{\bibfnamefont{J.}~\bibnamefont{Zhang}}, \bibinfo{author}{\bibfnamefont{S.}~\bibnamefont{Bhattacharjee}}, \bibnamefont{and} \bibinfo{author}{\bibfnamefont{B.}~\bibnamefont{Chakraborty}}, \bibinfo{journal}{Phys. Rev. Lett.} \textbf{\bibinfo{volume}{125}}, \bibinfo{pages}{118002} (\bibinfo{year}{2020}).

\bibitem[{\citenamefont{Nampoothiri et~al.}(2022)\citenamefont{Nampoothiri, D'Eon, Ramola, Chakraborty, and Bhattacharjee}}]{PhysRevE.106.065004}
\bibinfo{author}{\bibfnamefont{J.~N.} \bibnamefont{Nampoothiri}}, \bibinfo{author}{\bibfnamefont{M.}~\bibnamefont{D'Eon}}, \bibinfo{author}{\bibfnamefont{K.}~\bibnamefont{Ramola}}, \bibinfo{author}{\bibfnamefont{B.}~\bibnamefont{Chakraborty}}, \bibnamefont{and} \bibinfo{author}{\bibfnamefont{S.}~\bibnamefont{Bhattacharjee}}, \bibinfo{journal}{Phys. Rev. E} \textbf{\bibinfo{volume}{106}}, \bibinfo{pages}{065004} (\bibinfo{year}{2022}).

\bibitem[{\citenamefont{Liu et~al.}(2022)\citenamefont{Liu, Bojesen, Tabor, Mudie, Zaccone, Harrowell, and Petersen}}]{doi:10.1126/sciadv.abn0681}
\bibinfo{author}{\bibfnamefont{A.~C.~Y.} \bibnamefont{Liu}}, \bibinfo{author}{\bibfnamefont{E.~D.} \bibnamefont{Bojesen}}, \bibinfo{author}{\bibfnamefont{R.~F.} \bibnamefont{Tabor}}, \bibinfo{author}{\bibfnamefont{S.~T.} \bibnamefont{Mudie}}, \bibinfo{author}{\bibfnamefont{A.}~\bibnamefont{Zaccone}}, \bibinfo{author}{\bibfnamefont{P.}~\bibnamefont{Harrowell}}, \bibnamefont{and} \bibinfo{author}{\bibfnamefont{T.~C.} \bibnamefont{Petersen}}, \bibinfo{journal}{Science Advances} \textbf{\bibinfo{volume}{8}}, \bibinfo{pages}{eabn0681} (\bibinfo{year}{2022}).

\bibitem[{\citenamefont{Martinelli et~al.}(2023)\citenamefont{Martinelli, Caporaletti, Dallari, Sprung, Westermeier, Baldi, and Monaco}}]{PhysRevX.13.041031}
\bibinfo{author}{\bibfnamefont{A.}~\bibnamefont{Martinelli}}, \bibinfo{author}{\bibfnamefont{F.}~\bibnamefont{Caporaletti}}, \bibinfo{author}{\bibfnamefont{F.}~\bibnamefont{Dallari}}, \bibinfo{author}{\bibfnamefont{M.}~\bibnamefont{Sprung}}, \bibinfo{author}{\bibfnamefont{F.}~\bibnamefont{Westermeier}}, \bibinfo{author}{\bibfnamefont{G.}~\bibnamefont{Baldi}}, \bibnamefont{and} \bibinfo{author}{\bibfnamefont{G.}~\bibnamefont{Monaco}}, \bibinfo{journal}{Phys. Rev. X} \textbf{\bibinfo{volume}{13}}, \bibinfo{pages}{041031} (\bibinfo{year}{2023}).

\bibitem[{\citenamefont{Plimpton}(1995)}]{lammps}
\bibinfo{author}{\bibfnamefont{S.}~\bibnamefont{Plimpton}}, \bibinfo{journal}{J. Comp. Phys} \textbf{\bibinfo{volume}{117}}, \bibinfo{pages}{1} (\bibinfo{year}{1995}).

\end{thebibliography}

\end{document}